\documentclass[useAMS, usenatbib, usegraphicx]{mn2e}
\usepackage[fleqn, tbtags]{amsmath}
\usepackage{ifthen}
\usepackage{times}
\renewcommand{\vec}{\bmath}
\newcommand{\ud}{\rmn{d}}
\newcommand{\satellite}[1]{{\it{#1}}}
\newcommand{\invMpc}{\ \rmn{Mpc}^{-1}}
\newboolean{grey-scale}
\setboolean{grey-scale}{false}
\begin{document}
\title[Substructure: power spectrum and bispectrum]
{Effects of halo substructure on the power spectrum and bispectrum}
\author[Derek Dolney et al.]{
  Derek Dolney,\thanks{E-mail: dolney@astro.upenn.edu (DD);
                               bjain@physics.upenn.edu (BJ);
                               mtakada@hep.upenn.edu (MT)} 
  Bhuvnesh Jain,\footnotemark[1] 
  Masahiro Takada\footnotemark[1]\\
  Department of Physics and Astronomy, University of Pennsylvania,
  209 S 33RD ST FL 2, Philadelphia, PA 19104-6396, USA
}
\date{Accepted 2004 May 5. Received 2004 April 6; in original form
  2004 January 9}
\pagerange{\pageref{firstpage}--\pageref{lastpage}} \pubyear{2004}
\maketitle
\label{firstpage}
\begin{abstract}
  We study the effects of halo substructure and a distribution in the
  concentration parameter of haloes on large-scale structure statistics. 
  The effects on the power spectrum and bispectrum are studied
  on the smallest scales accessible from future surveys. We compare 
  halo-model predictions with results based on $N$-body simulations, 
  but also extend our predictions to 10-kpc scales which will
  be probed by future simulations.  
  We find that weak-lensing surveys proposed for the
  coming decade can probe the power spectrum on small enough scales
  to detect substructure in massive haloes. We discuss the prospects of
  constraining the mass fraction in substructure in view of partial 
  degeneracies with parameters such as the tilt and running of the
  primordial power spectrum.
\end{abstract}
\begin{keywords}
  galaxies: haloes -- cosmological parameters -- dark matter.
\end{keywords}
\section{Introduction}
The substructure of dark-matter haloes is a topic of interest in
cosmology, especially due to its relevance to the cold-dark-matter
model, which requires some degree of clumpiness to the matter
distribution. In a recent paper, \citet{KochanekDalal2003} argue that
substructure provides the best explanation (originally
proposed by \citealt{MaoSchneider1998}) of the anomalous
flux ratios found in gravitational lens systems. However, the observed
number of dwarf galaxy satellites in the Local Group is more than an
order of magnitude smaller than the number of subhaloes found in
numerical simulations (\citealt{KlypinEtAl1999, MooreEtAl1999a}). 

Observable statistics of
the matter distribution provide an approach to substructure studies
that is complementary to studies of strong-lens systems or direct
observations of satellite galaxies. Substructure effects are expected
to produce an observable effect in the power spectrum in the next
generation of lensing surveys. As we show in this paper, substructure
effects in the weak-lensing regime probe cluster scales primarily, as
opposed to strong-lensing systems, which probe galactic scales.

The halo model provides a description of non-linear gravitational
clustering which allows statistical quantities of interest to be
efficiently computed. The essential concepts of the halo model 
were used by \citet{NeymanScott1952} to describe the spatial
distribution of galaxies over fifty years ago. 
Their results were later generalized by \citet{ScherrerBertschinger1991}.
By combining the formalism of Scherrer \& Bertschinger with
results from $N$-body simulations such as a halo density profile (e.g.
that of \citealt*{NavarroFrenkWhite1997}) and a
halo mass function and bias parameters (e.g,
\citealt{ShethTormen1999}), one is able to calculate various
dark-matter and galaxy clustering statistics
(\citealt{MaFry2000, Seljak2000, PeacockSmith2000,
  ScoccimarroShethHuiJain2001, CoorayHu2001}).
See \citet{CooraySheth2002} for a comprehensive review of the halo
model.

Most work on the halo model approximates haloes as
spherically symmetric objects, described by a mass density which is a
smoothly-decreasing function only of the distance from the halo centre. 
However, the density distribution of haloes which form in $N$-body simulations
is not at all smooth (\citealt{NavarroFrenkWhite1997, MooreEtAl1999b}).
Rather, one finds that about 10 per cent of the mass is associated with
subhaloes (\citealt*{TormenDiaferioSyer1998, GhignaEtAl2000}). In the
large-scale structure literature to date, the effects of halo
substructure are typically assumed to be small for purposes of calculating
statistics of the dark-matter distribution.
This paper explores the validity of this assumption. We will make
quantitative statements
about the size of substructure effects on the 2D and 3D dark-matter power
spectra and bispectra. We will identify the length and
angular scales at which halo substructure becomes important.

The results of $N$-body simulations also suggest a distribution of halo
concentration parameters (\citealt{BullockEtAl2001, Jing2000}).
Like substructure, this effect is often ignored, though the formalism
to include it in the halo model is straightforward
(\citealt{CoorayHu2001}). This paper will quantify the effect of a
concentration parameter distribution on the power spectra.
\section{Theory} \label{sec:Theory}
To model substructure in a halo, the total mass $M$ of the halo is
divided into a smoothly distributed component
of mass $M_\rmn{s}$, and the remaining mass $m$ is placed in
subhaloes. We will use the terms subhalo and subclump
interchangeably. Ignoring substructure, the
halo model requires specification of the halo density
profile and the mass function, i.e. the number density of haloes of a
given mass. To include substructure,
one must specify not only the density profile of
the smooth component, the Fourier transform of which we denote
$M_\rmn{s}\, U(k;M_\rmn{s})$, but also the density profile of the subhaloes, with
Fourier transform $m\, u(k;m)$. The mass factors simply
provide the proper normalization; $U(k;M_\rmn{s})$ and $u(k;m)$ are
normalized to unity.
Suitable mass functions for the
number density both of parent haloes, $\ud N / \ud M$, and of
subhaloes, $\ud n / \ud m$, are also needed. In addition, the
distribution of subhaloes
within the parent halo must be specified. We denote its Fourier transform
by $U_\rmn{c}(k;M_\rmn{s})$. Using these ingredients,
\citet{ShethJain2003}
derive expressions for the power spectrum and the bispectrum. We
give some results from their paper that are relevant to this work in
the Appendix.

We use the mass function of \citet{ShethTormen1999} for the
number density of parent haloes of a given mass $M$:
\begin{equation} \label{eq:MassFunction}
  \ud M \, \frac{\ud N(M)}{\ud M}
        = \frac{\bar\rho}{M} \, A \, [1 + (a \nu)^{-p}]
          \, \sqrt{a \nu} \, e^{-a \nu / 2} \, \frac{\ud \nu}{\nu},
\end{equation}
where $\bar \rho$ denotes the average density of the Universe at the
present epoch, and
\begin{equation}
  \nu(M, z) := \left[ \frac{\delta_\rmn{c}(z)}{D(z) \sigma(M)} \right]^2
\end{equation}
is the peak height defined in terms of $\delta_\rmn{c}(z)$, the threshold
overdensity in the spherical-collapse model, $D(z)$, the growth
factor for density perturbations, and $\sigma(M)$, the rms fluctuations in
the matter density field at the present epoch smoothed with a top-hat
window of radius $(3 M / 4 \upi \bar\rho)^{1 / 3}$. The normalization
constant is fixed by requiring that all mass be contained within a
parent halo, and turns out to be $A = 0.116$. For $\delta_\rmn{c}$,
we use the fitting formula of \citet{Henry2000}. We use haloes with
mean interior density 180 times the background
density and take $a = 0.67$ and $p = 0.33$, so that our mass function is
universal (\citealt{White2002}). The notation of this paper reflects our
choice of halo boundary: $r := r_{180b}$, $M := M_{180b}$ and
$c := r_{180b} / r_\rmn{s}$, where $r_\rmn{s}$ is the usual scale radius
parameter of the halo density profile.

We use the density profile of \citet{NavarroFrenkWhite1997}, hereafter
NFW, for both the smooth component and the
subhaloes. It's Fourier transform is most efficient to compute using
sine and cosine integrals:
\begin{equation}
  U(k, m) = \cos\kappa
            \left\{\rmn{Ci}\left[\kappa(1 + c)\right] 
               - \rmn{Ci}(\kappa)\right\}
              - \frac{\sin{\kappa c}}{\kappa(1 + c)},
\end{equation}
with $\kappa := k r_{180b} / c$,
\begin{equation}
  \rmn{Si}(x) := \int_0^x \ud r \, \frac{\sin r}{r}
\end{equation}
and
\begin{equation}
  \rmn{Ci}(x) := -\int_x^\infty \ud r \, \frac{\cos r}{r}.
\end{equation}
We assume the distribution of subhaloes within
a parent halo follows the mass distribution of the smooth
component. That is, $U_\rmn{c}(k; M_\rmn{s})$ is also given by the NFW profile,
normalized to unity. Simulations suggest that the
abundance of subhaloes in the parent halo traces the density profile
of the parent quite well except in the most dense region near its
centre where the
abundance of subhaloes is lessened due to tidal disruptions. The
subhalo abundance in the inner region is not yet well resolved in
simulations. \citet*{ChenKravtsovKeeton2003} propose a modified NFW
profile that is constant or zero inside of a core radius which
they take to be 10--20 kpc for a $10^{12}$-$M_\odot$ halo. For a halo of
this mass, only 4 per cent of its matter lies within the core radius. Hence,
were we to use the latter profile, the effect would be to move only
4 per cent of our subhaloes further from the centre. We will comment briefly
on the effect we expect this to have on the power spectrum in the
Discussion.

We use the concentration parameters found by
\citet{BullockEtAl2001}, for all three profiles at redshift $z$, which
are
\begin{equation} \label{eq:BullockSmooth}
  \bar c(M, z) = 9 \, (1 + z)^{-1} 
                 \left[ \frac{M_\rmn{vir}(M)}{M_*} \right]^{-0.13}
                 \quad \rmn{(smooth)},
\end{equation}
for the smooth component of a halo and
\begin{equation} \label{eq:BullockSubhalo}
  \bar c(m, z) = 7.5 \, (1 + z)^{-1}
                 \left[ \frac{m_\rmn{vir}(m)}{M_*} \right]^{-0.30}
                 \quad \rmn{(subhalo)},
\end{equation}
for all subhaloes. The non-linear mass scale $M_*$ is defined by the
relation $\nu(M_*, 0) = 1$.
The translation $M_\rmn{vir}(M)$ between different halo
mass definitions is derived in the appendix of
\citet{HuKravtsov2003}. Bullock et al.~find that the concentration
parameter for subhaloes has some dependence on the local density around
the subhalo. We ignore this effect here, as it is difficult to include
in our model, and we do not believe inclusion would seriously affect
our conclusions. \citet{HuffenbergerSeljak2003} find a different
concentration
parameter by $\chi^2$ minimization against the non-linear fitting
formula of \citet{SmithEtAl2003}:
\begin{equation}
  \bar c(M, z) = 11 \, (1 + z)^{-1}
                 \left( \frac{M}{M_*} \right)^{-0.05}.
\end{equation}

Numerical simulations suggest a log-normal distribution for the
concentration parameter about the mean value $\bar c$,
\begin{equation} \label{eq:cDist}
  p(c;m,z) \, \ud c = \frac{\ud (\ln c)}{\sqrt{2 \upi \sigma_{\ln c}^2}}
                   \exp \left\{ {-\frac{\ln^2[c / \bar c(m, z)]}
                                {2 \sigma_{\ln c}^2}}
                        \right\},
\end{equation}
with a width $\sigma_{\ln c} = 0.3$
(\citealt{Jing2000, BullockEtAl2001}).
Unless otherwise noted, we use the mean values given in
equations (\ref{eq:BullockSmooth}) and (\ref{eq:BullockSubhalo}).

We assume a biasing prescription for the clustering of parent
haloes. That is, the halo power spectrum is given 
in terms of the linear mass power spectrum $P_L(k)$ as
\begin{equation}
  P(k, z; M_1, M_2) = b_1(\nu_1) \, b_1(\nu_2) P_L(k, z),
\end{equation}
where $\nu_i := \nu(M_i, z)$.
We use the bias parameter of \citet{ShethTormen1999}:
\begin{equation}
  b_1(\nu) = 1 + \frac{a \nu - 1}{\delta_\rmn{c}} 
               + \frac{2 p / \delta_\rmn{c}}{1 + (a \nu)^p}.
\end{equation}

For the subhaloes, we assume a power-law mass function
suggested by numerical simulations
(\citealt{TormenDiaferioSyer1998, GhignaEtAl2000}):
\begin{equation} \label{eq:powerLaw}
  \frac{\ud n(m; M)}{\ud m}\, \ud m =
    N_0\left(\frac{M}{m}\right)^\mu\frac{\ud m}{m},
\end{equation}
where $\mu < 1$. The normalization $N_0$ is determined 
by the mass fraction contained in subhaloes, $f$:
\begin{equation} \label{eq:subNorm}
  f := \frac{M - M_\rmn{s}}{M} = \int \ud m \, \frac{m}{M}
       \frac{\ud n(m; M)}{\ud m}.
\end{equation}
Numerical simulations suggest $\mu \approx 0.9$ and
$f \approx 10$ per cent. We use these values unless otherwise noted. In
addition, simulations suggest that subhaloes heavier than 1 per cent of the
total mass in the parent halo are rare
(\citealt{TormenDiaferioSyer1998}; \citealt{ChenKravtsovKeeton2003}), so we
only include subhaloes with masses $m \leq 0.01 M$.

We calculate the weak-lensing convergence using Limber's
approximation (\citealt{Limber1953, Kaiser1992}), so that
the convergence power spectrum and bispectrum are given by
\begin{equation}
  P^\kappa(l)=\int_0^{\chi_\rmn{s}} \ud \chi\,
  \frac{W^2(\chi)}{d_\rmn{A}^2(\chi)}
                    \, P\left(\frac{l}{d_\rmn{A}(\chi)}\right)
\end{equation}
and
\begin{equation}
  B^\kappa(l_1,l_2,l_3) =
    \int_0^{\chi_\rmn{s}} \ud \chi \,
    \frac{W^3(\chi)}{d_\rmn{A}^4(\chi)} \,
    B \left(\frac{l_1}{d_\rmn{A}(\chi)}, \frac{l_2}{d_\rmn{A}(\chi)},
            \frac{l_3}{d_\rmn{A}(\chi)} \right).
\end{equation}
In the next section we will also use the reduced bispectrum, defined
as $B/3P^2$ for equilateral triangles.

We use the standard weak-lensing weight function (see, e.g.
\citealt{BartelmannSchneider2001})
\begin{equation}
  W(\chi) = \frac{3}{2} \Omega_\rmn{m} H_0^2 a^{-1}
            \frac{d_\rmn{A}(\chi) \, d_\rmn{A}(\chi_\rmn{s} - \chi)}{d_\rmn{A}(\chi_\rmn{s})},
\end{equation}
where $\chi$ is the comoving distance, $\chi_\rmn{s}$ is the comoving
distance of the source galaxies, $d_\rmn{A}(\chi)$ is the
comoving angular diameter distance and $H_0$ is the present-day
value of the Hubble parameter. For simplicity, we place all
source galaxies at redshift 1.

We will calculate the covariance of the convergence field by adopting
a binning scheme. Let $P^\kappa_i$ denote the power averaged over a
bin of width $\Delta_i$ centred on the mode $l_i$. Our bins are
chosen to have width $\Delta_i / l_i = 0.3$. In this notation, the
covariance can be expressed as
(\citealt{CooraySheth2002, MeiksinWhite1999})

\begin{equation} \label{eq:covariance}
  C_{ij} = \frac{1}{4 \upi f_\rmn{sky}}
    \Bigg\{\frac{4 \upi}{l_i \Delta_i}
      \left[P^\kappa_i 
        + \frac{\sigma_{\gamma_x}^2}
               {\bar n}\right]^2 \delta_{ij}
    + \ T^\kappa(\vec l_i, -\vec l_i, \vec l_j, -\vec l_j) \Bigg\},
\end{equation}
where $f_\mathrm{sky}$ is the fraction of the sky surveyed,
$\sigma_{\gamma_x}$ is the per component rms intrinsic ellipticity
of the sample galaxies, $\bar n$ is the average galaxy number density
in the survey. The first term is the sample variance; the second the
shot noise. The last term, $T^\kappa$, is the trispectrum
contribution, arising from non-Gaussianity in the convergence
field.

It is likely that a space-based deep imaging survey will offer the best
prospects for probing 
the small scales where substructure effects are important. 
We adopt parameters expected
for the \satellite{SNAP} satellite lensing survey (e.g. \citealt{MasseyEtAl2003}):
$\bar n = 100 \ \rmn{gal} / \rmn{arcmin}^2$,
$f_\rmn{sky} = 2$ per cent, $\sigma_{\gamma_x} = 0.34 / \sqrt 2 = 0.24$.

We use the covariance of the power spectrum 
to perform a least-squares fitting to a fiducial
model $\bar P^\kappa$ by calculating $\chi^2$,
\begin{equation} \label{eq:chi2}
  \chi^2 = \sum_{ij}(P_i^\kappa - \bar P_i^\kappa)C_{ij}^{-1}
     (P_j^\kappa - \bar P_j^\kappa),
\end{equation}
and identifying $1\sigma$, $2\sigma$ and $3\sigma$ contours as those
with $\Delta \chi^2 = $ 2.30, 6.17 and 11.8, respectively, which is
appropriate for the two-parameter fits we will use.

We use a $\Lambda \rmn{CDM}$ cosmological model with parameters
$\Omega_\rmn{m} = 0.3$, $\Omega_\Lambda = 0.7$, $\Omega_\rmn{b} = 0.05$,
$h = 0.7$, $\sigma_8 = 0.9$ and the BBKS transfer function
(\citealt{BardeenEtAl1986}) with the shape parameter of
\citet{Sugiyama1995}.
\section{Results} \label{sec:results}
\begin{figure}
  \ifthenelse{\boolean{grey-scale}}
             {\includegraphics[width = 84mm]{fig1_grey}}
             {\includegraphics[width = 84mm]{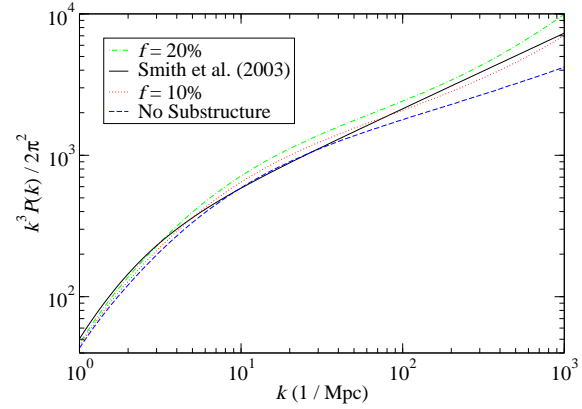}}
  \caption{Halo-model calculations of the dimensionless power spectrum
    with different
    fractions of the halo mass placed in NFW subhaloes.}
  \label{fig:fractions}
\end{figure}

Fig.~\ref{fig:fractions} compares halo-model calculations of the 
three-dimensional power spectrum with
the non-linear fitting formula of \citet{SmithEtAl2003}.
In this figure, different fractions
of the halo mass are placed in substructure. Adding substructure
can be seen to increase power at scales $k \ga 1 \invMpc$.
Substructure models predict more
power than Smith et al.~for $10 \la k \la 100 \invMpc$, but
the model with 10 per cent of the halo mass in substructure shows
better agreement at smaller scales (though it
should be noted that Smith et al.~has been extrapolated beyond
$k \ga 40 \, h \invMpc$).
The power spectrum is fairly insensitive to the exponent of the
subhalo mass function, $\mu$. Reducing $\mu$ from 0.9 to 0.7 can
increase the power spectrum by about 4 per cent at length scales
$k \ga 1 \invMpc$.

\begin{figure}
  \ifthenelse{\boolean{grey-scale}}
             {\includegraphics[width = 84mm]{fig2_grey}}
             {\includegraphics[width = 84mm]{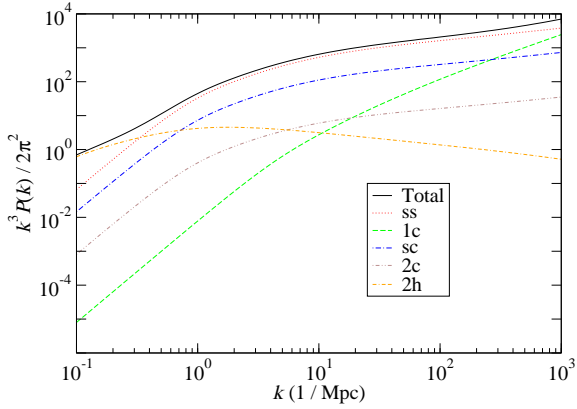}}
  \caption{The dimensionless power spectrum separated into
    the contributions from each of the terms in
    equations (\ref{eq:P_ss})--(\ref{eq:P_2h}). The order of the lines in
    the legend matches the order at $k = 10^3 \invMpc$.}
  \label{fig:contrib}
\end{figure}

In Fig.~\ref{fig:contrib} we divide the power spectrum into
contributions from the various terms in
equations (\ref{eq:P_ss})--(\ref{eq:P_2h}). On scales $k < 200 \invMpc$,
the substructure contribution to
the power spectrum is dominated by the density correlation between the smooth
component of the parent halo and a subclump, denoted $P_\rmn{sc}$.
The correlation of the density in a single subclump, $P_\rmn{1c}$, or
between two different subclumps, $P_\rmn{2c}$, is negligible for
parameter values of interest
except for very high wavenumbers (where baryonic effects are likely
to play a role). One should also note that the power of the smooth
component only, $P_\rmn{ss}$, still dominates all other contributions
on all scales considered, except for the largest scales where the two-halo
term, $P_\rmn{2h}$, models the linear power spectrum.

\begin{figure}
  \ifthenelse{\boolean{grey-scale}}
             {\includegraphics[width = 84mm]{fig3_grey}}
             {\includegraphics[width = 84mm]{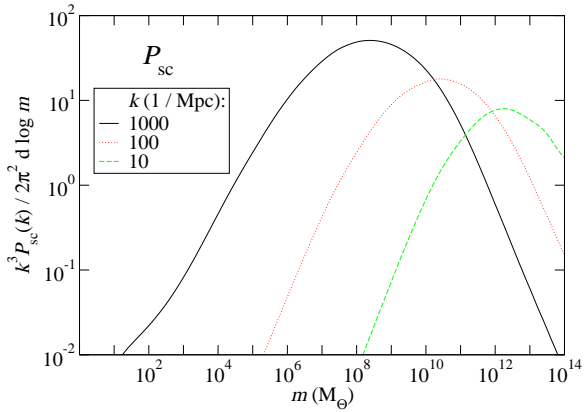}}
  \caption{Contributions from the smooth-subclump term to
    the power spectrum per logarithmic mass bin.
    Integrating these curves against $\ud \log m$
    gives $P_\rmn{sc}(k)$.}
 \label{fig:contribMass}
\end{figure}

In Fig.~\ref{fig:contribMass}, we leave the
integral over subhalo mass uncalculated for the $P_\rmn{1c}$ and
$P_\rmn{sc}$ terms, so that one can see which
subhalo masses provide the most power.
The figures suggest that most of the extra power due to substructure
between $10 \le k \le 100 \invMpc$, comes from subhaloes of mass
$10^{10} \la m \la 10^{12} M_\odot$. Subhaloes of this mass are
likely contained in parent haloes of mass
$\ga 10^{12}$--$10^{14} M_\odot$.
Finally, since the weak-lensing efficiency peaks for
$z \approx 0.3$--$0.5$, the conclusion is that the weak-lensing 
power spectrum can probe intermediate-redshift group and cluster 
mass haloes (provided the signal-to-noise permits its measurement
as discussed below). 
This should make weak lensing as a probe of substructure a good
complement to studies of strong lensing of quasars, which probe
substructure in galactic haloes. 

\begin{figure}
  \ifthenelse{\boolean{grey-scale}}
             {\includegraphics[width = 84mm]{fig4_grey}}
             {\includegraphics[width = 84mm]{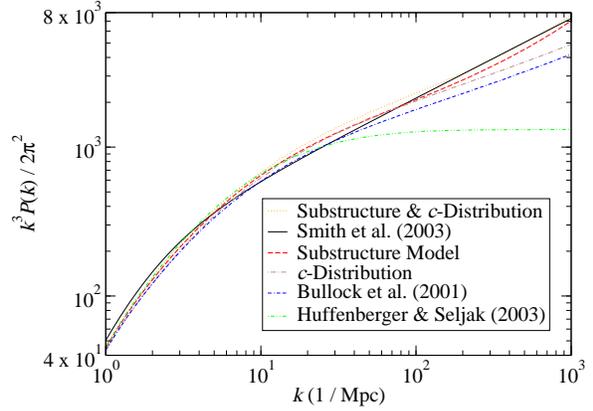}}
  \caption{Comparison of dark-matter power spectra predicted by
    various models, including substructure, a log-normal distribution
    of halo concentration parameters as in equation (\ref{eq:cDist}), and the
    non-linear fitting formula of Smith et al. The order of the lines in
    the legend matches the order at $k = 10^3 \invMpc$.}
 \label{fig:PS_models}
\end{figure}

In Fig.~\ref{fig:PS_models} we plot a number of power spectra
calculated with various models. Henceforth, curves titled
`Bullock et al.~(2001)' and `Huffenberger \& Seljak (2003)' refer
to halo-model calculations without substructure, using the
concentration parameters found in those references.
It should be noted that Huffenberger \& Seljak
obtained their concentration parameter by fitting to
\citet{SmithEtAl2003} only up $k = 40 \, h \invMpc$. Both
Huffenberger \& Seljak and Smith et al.~have been extrapolated beyond
$k = 40 \, h \invMpc$, and
the apparently poor agreement of their model with Smith et
al.~beyond this wavelength should be attributed to this extrapolation.

It is apparent that adding either substructure or the
concentration parameter distribution affects power only on scales
$k \ga 10 \invMpc$.
Substructure provides a 10--60 per cent increase in power on the
scales $10 \le k \le 1000 \invMpc$, with the largest factor
corresponding to the smallest scales. The $c$-distribution provides
only a 6--10 per cent increase in power over the same scales.

It is interesting to note that while the substructure model increases
power relative to \citet{SmithEtAl2003} or \citet{BullockEtAl2001},
the concentration parameter of \citet{HuffenbergerSeljak2003}
decreases the relative power. And so, while all models may be
consistent with Smith et al.~at scales $k \le 10 \invMpc$,
the better model should reveal itself as smaller scales are probed in
higher-resolution simulations and observations. If the halo model is
an appropriate physical description of mass clustering, then the 
substructure model presented here should provide the most robust
predictions for the power spectrum on small scales.

\begin{figure}
  \ifthenelse{\boolean{grey-scale}}
             {\includegraphics[width = 84mm]{fig5_grey}}
             {\includegraphics[width = 84mm]{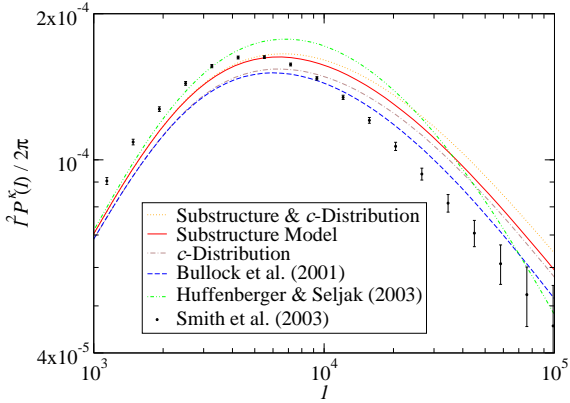}}
  \caption{Various model predictions of dimensionless convergence
    power spectrum. Compares the substructure model and
    $c$-distribution to non-substructure models. The error bars are
    the diagonal part of the covariance calculated with
    equation (\ref{eq:covariance}) assuming
    $\bar n = 100$ gal/arcmin$^2$,
    $f_\rmn{sky} = 2$ per cent and
    $\langle \tilde\gamma_\rmn{int}^2 \rangle^{1 / 2} = 0.34$.
    The trispectrum contribution to equation (\ref{eq:covariance}) is
    approximated by a halo-model calculation. The order of the lines in
    the legend matches the order at $l = 10^5$.}
 \label{fig:P_kappa}
\end{figure}

\begin{figure}
  \ifthenelse{\boolean{grey-scale}}
             {\includegraphics[width = 84mm]{fig6_grey}}
             {\includegraphics[width = 84mm]{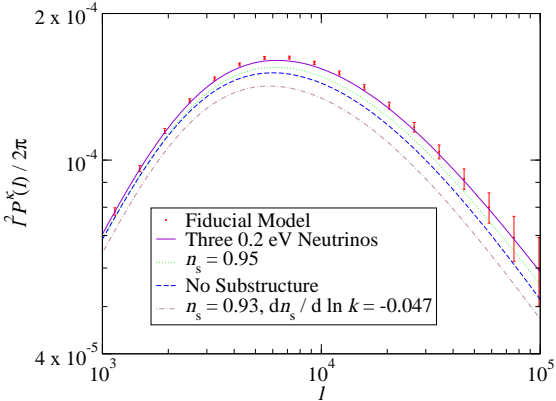}}
  \caption{Dimensionless convergence power spectra for models with
    other phenomena that produce effects similar to substructure.
    The error bars are the diagonal part of the
    covariance calculated with equation (\ref{eq:covariance}) 
        as in Fig.~\ref{fig:P_kappa}.  The order of the lines in
    the legend matches the order at $l = 10^5$.}
  \label{fig:P_kappaModels}
\end{figure}

In Fig.~\ref{fig:P_kappa} we show the dimensionless convergence power
spectrum predictions for the various models, assuming source galaxies
at redshift 1. Again, $c$-distribution
and substructure each provide about a 10 per cent increase in power at small
scales, though the effect of substructure gets larger for the smallest
scales. The halo-model calculations are discrepant with 
\citet{SmithEtAl2003} at the 10--20 per cent level. Most of the
contribution to $P^\kappa(l)$ for $10^3 \le l \le 10^5$
comes from $P(k)$ with $1 \le k \le 100 \invMpc$. At these scales,
the transition between the one-halo and two-halo terms is
important. \citet{ZehaviEtAl2003} have proposed a scale-dependent
correction to the bias parameter and a method to account for halo
exclusion that may improve the halo model on intermediate
scales. \citet{TakadaJain2003} employ an approximate correction for
halo exclusion to obtain a more well-behaved three-point correlation
function on scales $\approx 1 \ \rmn{Mpc}$. Given the error bars 
expected from future surveys, it is clear that improved
precision in simulations as well as model predictions is needed. Some
additional cosmological models are plotted in
Fig.~\ref{fig:P_kappaModels}, which we will comment more on shortly.

\begin{figure}
  \ifthenelse{\boolean{grey-scale}}
             {\includegraphics[width = 84mm]{fig7a_grey}
              \includegraphics[width = 84mm]{fig7b_grey}}
             {\includegraphics[width = 84mm]{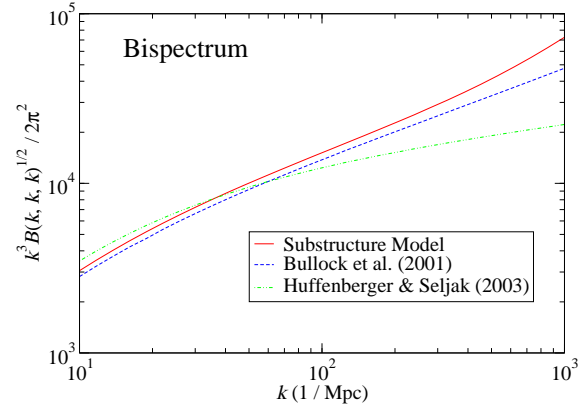}
              \includegraphics[width = 84mm]{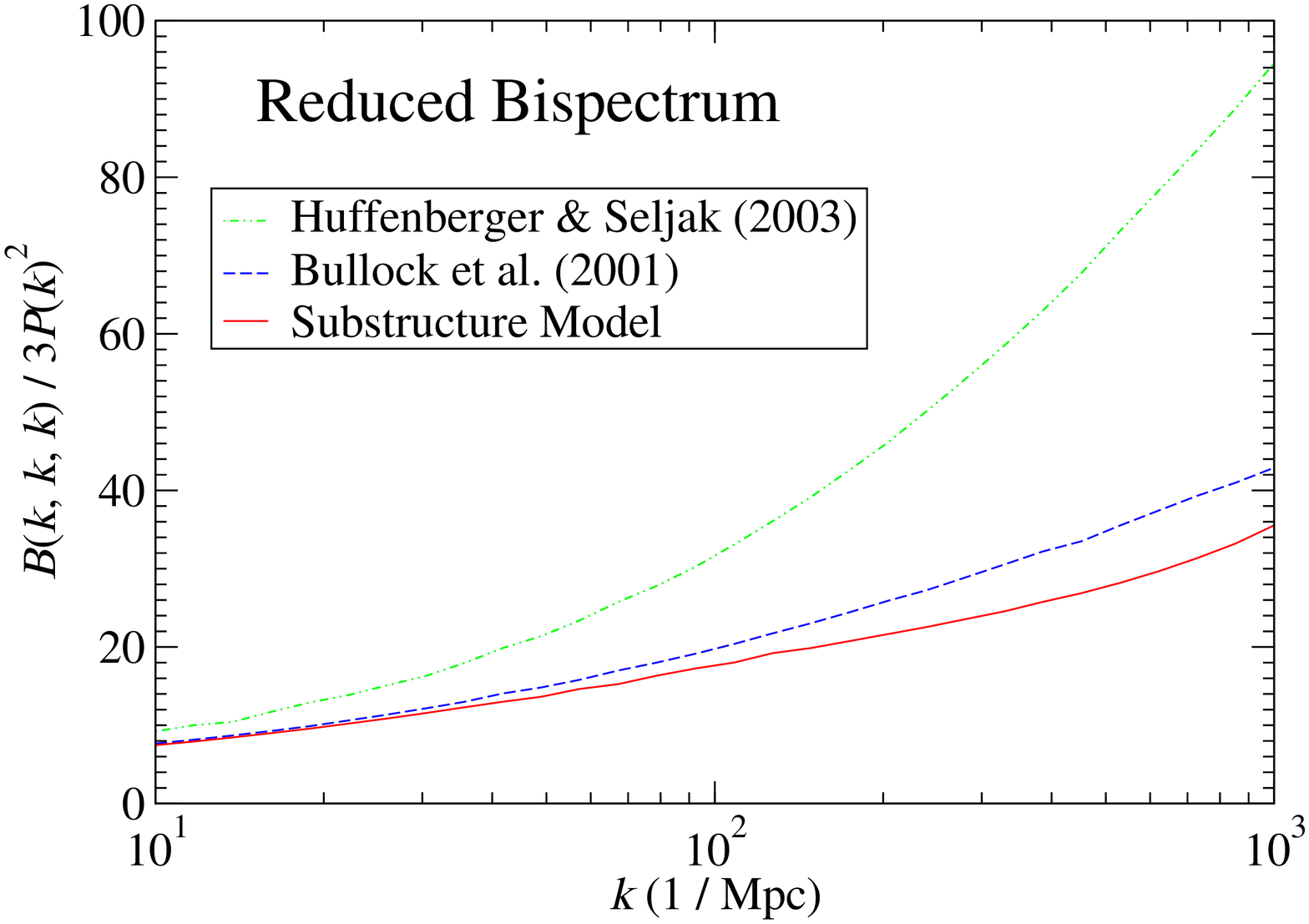}}
  \caption{Various model predictions of the equilateral bispectrum. The
    upper panel shows the dimensionless equilateral bispectrum; the
    lower panel is the equilateral reduced bispectrum.}
  \label{fig:Q_eq}
\end{figure}

\begin{figure}
  \ifthenelse{\boolean{grey-scale}}
             {\includegraphics[width = 84mm]{fig8a_grey}
              \includegraphics[width = 84mm]{fig8b_grey}}
             {\includegraphics[width = 84mm]{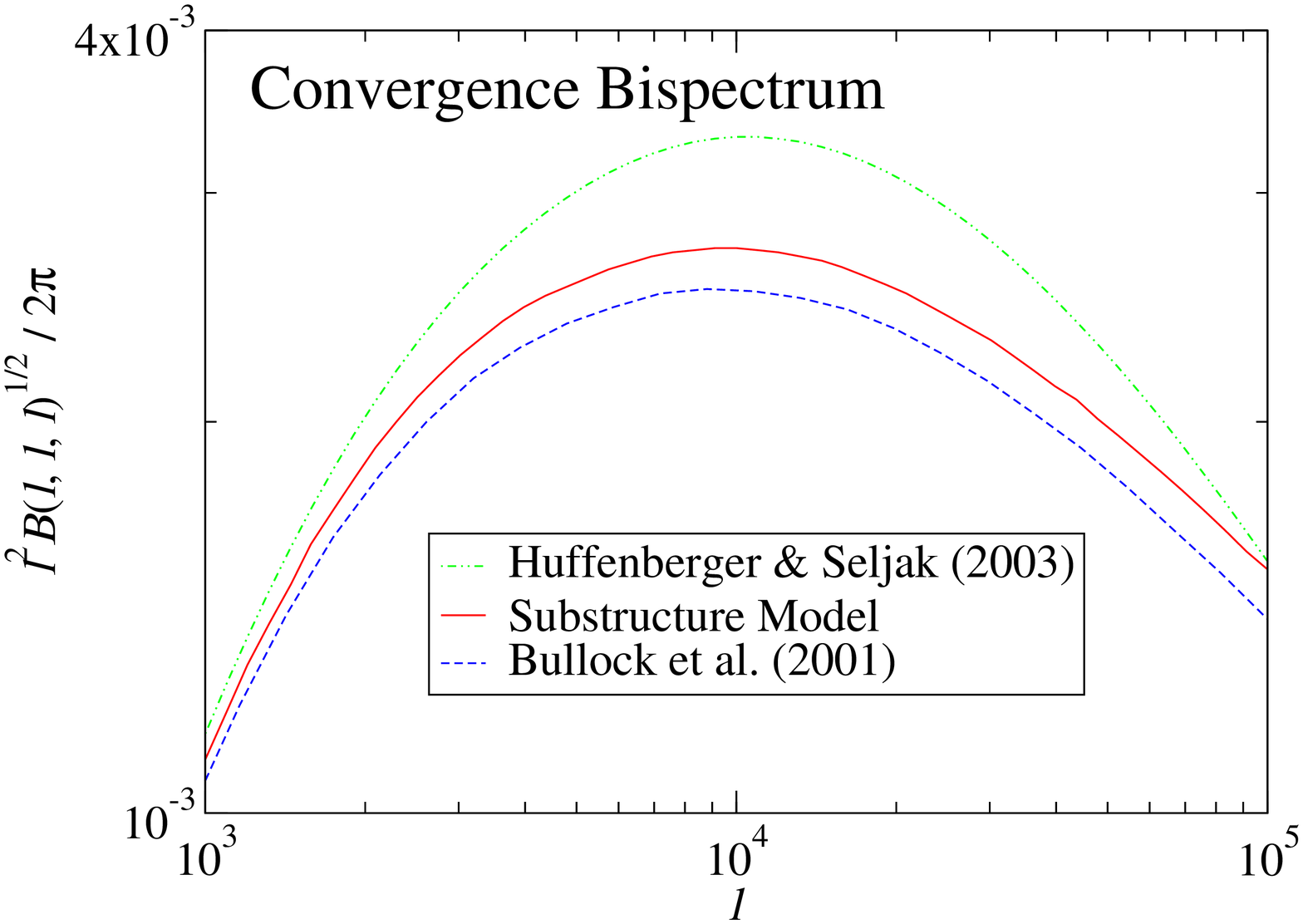}
              \includegraphics[width = 84mm]{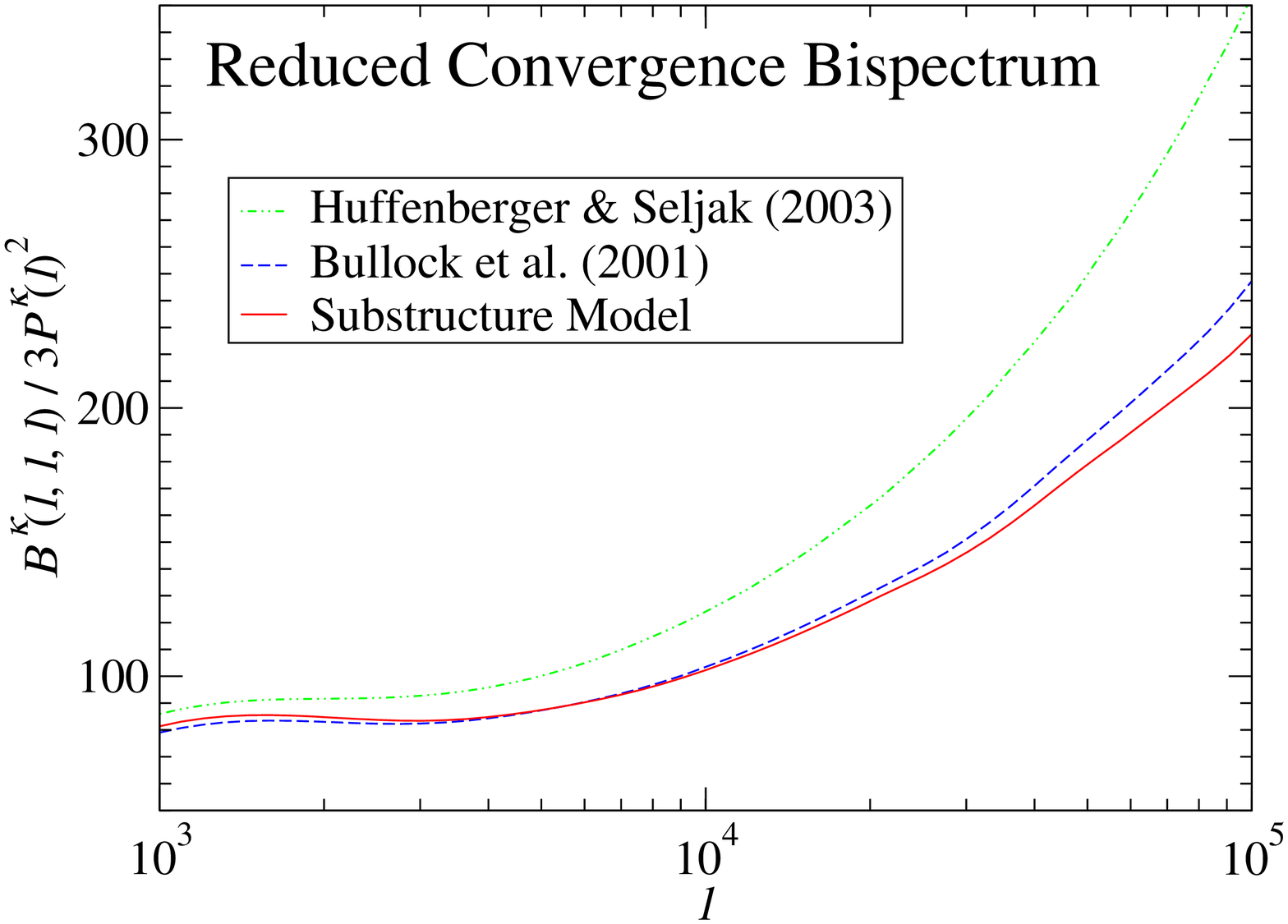}}
  \caption{Various model predictions of the equilateral convergence
    bispectrum. Only the one-halo term is plotted. The upper panel shows
    the dimensionless equilateral convergence bispectrum; the lower
    panel is the equilateral reduced convergence bispectrum.}
  \label{fig:Q_kappa}
\end{figure}

In Figs.~\ref{fig:Q_eq} and \ref{fig:Q_kappa}, we plot the predicted
three-dimensional bispectrum and convergence bispectrum. 
It is interesting to note
that, though substructure increases the bispectra, it causes a
decrease in the reduced bispectra. That is, the substructure
contribution to the power spectrum factor in the denominator exceeds
its contribution to the numerator in the reduced bispectrum. This is
due to the fact that the dominant substructure terms are the smooth-subclump
for the power spectrum and the smooth-smooth-subclump term for the
bispectrum.
The $P_\rmn{sc}$ term scales as $(1 - f)f$, while the $B_\rmn{ssc}$
term scales as $(1 - f)^2 f$,
so the reduced bispectrum scales as $1/f$
over most scales. Thus by using both the bispectrum and power
spectrum, degeneracies between the substructure fraction and other 
cosmological parameters can be broken. 
We discuss this further in Section \ref{sec:discussion}.

\begin{figure}
  \ifthenelse{\boolean{grey-scale}}
             {\includegraphics[width = 84mm]{fig9_grey}}
             {\includegraphics[width = 84mm]{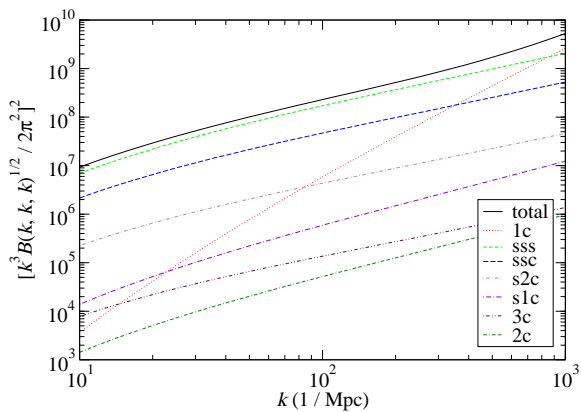}}
  \caption{The one-halo term of the equilateral bispectrum
    separated into the contributions
    from each of the terms in equations (\ref{eq:B_sss})--(\ref{eq:B_3c}). The
    order of the lines in
    the legend matches the order at $k = 10^3 \invMpc$.}
  \label{fig:Bcontrib}
\end{figure}

Shown in Fig.~\ref{fig:Bcontrib} is the substructure
contribution to the bispectrum separated into each of the terms in
equations (\ref{eq:B_sss})--(\ref{eq:B_3c}). On the scales of interest, the
substructure contributions to the bispectrum are dominated by
the smooth-smooth-subclump term,
$B_\rmn{ssc}$, and the bispectrum for individual subclumps,
$B_\rmn{1c}$. The meaning of these terms is discussed more completely
in the Appendix. It is interesting that the bispectrum contribution
from single subhaloes, $B_\rmn{1c}$, provides a larger contribution
than the smooth component alone, $B_\rmn{sss}$, for
$k \approx 10^3 \invMpc$, but on these scales a purely
gravitational calculation is not likely to be valid.


\begin{figure}
  \ifthenelse{\boolean{grey-scale}}
             {\includegraphics[width = 80mm]{fig10a_grey}
              \includegraphics[width = 80mm]{fig10b_grey}}
             {\includegraphics[width = 80mm]{fig10a}
              \includegraphics[width = 80mm]{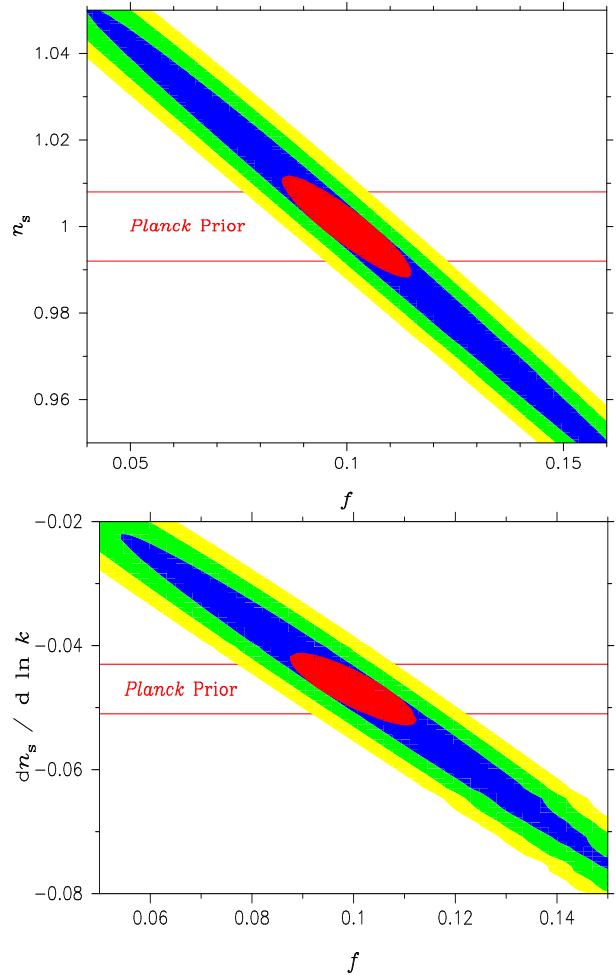}}
  \caption{Contour plots on two-dimensional planes through
    $\chi^2$-space (no marginalization). In the upper panel, the
    spectral index and substructure mass fraction vary with no running
    spectral index; in the lower  panel, the spectral
    index is fixed at $n_\rmn{s} = 0.93$ and a running parameter is varied along
    with the substructure mass fraction. The innermost contour of each
    plot is a one-sigma contour where we have applied prior constraints
    as expected from the \satellite{Planck} mission: $\sigma_{n_\rmn{s}} = 0.008$ and
    $\sigma_{\ud n_\rmn{s} / \ud \ln k} = 0.004$
    (\citealt{EisensteinHuTegmark1998}). The other contours are one-,
    two- and three-sigma with no priors.}
  \label{fig:chi2}
\end{figure}

Results from the \satellite{WMAP} satellite have constrained the tilt
and run of the
primordial power spectrum. Tilt affects the small scale power in a way
similar to substructure (cf. Fig.~\ref{fig:P_kappaModels}), so
we consider next the feasibility of weak lensing to
separate the effects of substructure from tilt
and run in the primordial power spectrum. The results of examining the
three parameter $\chi^2(f, n_\rmn{s}, \ud n_\rmn{s} / \ud \ln k)$ are shown in
Fig.~\ref{fig:chi2}. We
use the $f = 10$ per cent substructure model as the fiducial model, and
calculate the $\chi^2$ for $P^\kappa(l)$
relative to the fiducial model using modes in the range
$10^3 \le l \le 10^5$. The covariance matrix used is that in
equation (\ref{eq:covariance}). In the
upper panel, we assume no running and use $n_\rmn{s} = 1.00$ for the fiducial
model. In the lower panel, we fix $n_\rmn{s} = 0.93$ and allow a running
spectral index, as described in \citet{SpergelEtAl2003}, to vary about
a fiducial value of $\ud n_\rmn{s} / \ud \ln k = -0.047$
(\citealt{SpergelEtAl2003}). Note that
we do not marginalize over the third parameter in these figures. We
also apply priors as expected from the upcoming \satellite{Planck} mission:
$\sigma_{n_\rmn{s}} = 0.008$ and $\sigma_{\ud n_\rmn{s} / \ud \ln k} = 0.004$
(\citealt*{EisensteinHuTegmark1998}).
These figures suggest that with cosmic microwave background (CMB)
priors, future surveys that can measure 
$P^\kappa$ down to 0.1-arcmin scales
can constrain the substructure mass fraction to about $\pm 2$ per cent. 
\section{Discussion} \label{sec:discussion}
We have shown that on small scales, the contribution of CDM
substructure is important in accurate predictions of the power 
spectrum. In particular, it may help resolve the small scale discrepancy 
between $N$-body simulation results and
the halo model. Our model with 10 per cent substructure and a concentration
parameter distribution fits the small scale power spectrum
extrapolated from \citet{SmithEtAl2003} quite well. 
Higher-resolution $N$-body simulations are needed to test our model.

We agree with \citet{HuffenbergerSeljak2003} that a concentration
parameter distribution alone cannot resolve the discrepancy between
the halo model and \citet{SmithEtAl2003}. We have shown, however, that
the distribution increases power on small scales by about 10 per cent, which is
comparable to the accuracy of the Smith et al.~formula. 
Hence the scatter in the concentration parameter needs to be taken into
account on scales smaller than $k = 10 \invMpc$. 

Our examination of the prospects for parameter constraints are by no
means complete. A more careful treatment must marginalize over a 
larger parameter space $\chi^2(\Omega_\rmn{m}, \Omega_\rmn{b}, \Omega_\nu,
\sigma_8, n_\rmn{s}, \ud n_\rmn{s} / \ud \ln k, f,...)$, and
account for survey limitations such as intrinsic alignment of source
galaxies, the limits imposed by galaxy deblending issues
and B-mode contamination. Most 
parameters, like $\Omega_\rmn{m}$ and $\sigma_8$, are not very degenerate
with substructure, since they affect the power spectrum and bispectrum
on all scales, while substructure modifies the statistics only on small
scales. Such parameters can be well constrained from large
scale information and probably not confuse a measurement of
$f$. Moreover, other observations constrain
$\Omega_\rmn{m}$ and $\sigma_8$, such as the cosmic microwave background,
supernovae observations, galaxy clusters and
galaxy clustering statistics.

The situation is similar for neutrino mass and baryon fraction,
though to a somewhat lesser
extent. While these parameters affect the lensing power spectrum
they also produce some suppression on 
scales with $l < 10^4$ where substructure has little effect. 
This will help us to constrain $\Omega_\nu$, $\Omega_\rmn{b}$
and $f$ separately. The \satellite{WMAP} results suggest that neutrinos are less
massive than 0.2 eV (\citealt{SpergelEtAl2003}). We have shown that the
power spectrum is much less sensitive to such neutrinos than
substructure at the 10 per cent level (Fig.~\ref{fig:P_kappaModels}). 
With regard to baryons, their abundance is
well constrained from CMB data and big-bang
nucleosynthesis. Future observations are expected to reveal baryon
oscillations in the weak-lensing signal which will allow for even
tighter constraints on the baryon abundance.

The effects of the parameters $c_0$, $\beta$ and $\alpha$, which
denote the normalization and slope of the halo concentration and the
inner slope of the halo profile, can be
separated from substructure effects with information from the reduced
bispectrum. As we noted in Sec.~\ref{sec:results}, substructure
produces a proportionately larger increase in the power spectrum than
in the equilateral bispectrum. The result is that substructure
amplifies the power spectrum, but attenuates the reduced
bispectrum. This differs from the behavior of $c_0$, $\alpha$ and
$\beta$, as demonstrated by \citet{TakadaJain2003}, who find that
these two parameters change the power spectrum and reduced bispectrum
in the same sense.

Of the parameters considered, the tilt and run of the primordial power
spectrum appear to produce an effect most similar to halo substructure.
Our analysis in the previous section shows that the next generation of
weak-lensing surveys will likely provide interesting $n_\rmn{s} - f$ and
$\ud n_\rmn{s} / \ud \ln k - f$ constraints. Applying priors such as results from
the \satellite{Planck} mission to these constraints should allow a
detection of substructure. We find that
a substructure mass fraction constraint at the $\pm 2$ per cent
is possible if observational errors can be controlled. 

Numerical simulations indicate that, as a result of tidal disruptions
in particular, a modified NFW profile for the distribution of
subhaloes in the parent halo may be more appropriate. As we have already
pointed out, simulations have not yet provided us with a
definitive substructure-abundance profile for the centre-most regions of
halos. Even taking the abundance to be zero within a core radius of
order 10 kpc, as in \citet{ChenKravtsovKeeton2003}, we would
redistribute only 4 per cent of our subhaloes. So while the magnitude of the
$P_\rmn{sc}$ term may be decreased by this order on $100$-$\rmn{Mpc}^{-1}$
scales, smaller scales are dominated by the $P_\rmn{1c}$ contribution
(cf. Fig.~\ref{fig:contrib}), which does not depend on the subhalo
distribution.

We thank Gary Bernstein and Ravi Sheth for helpful discussions. This
work is supported in part by NASA through grant NAG5-10924 and
NSF through grant AST03-07297. 
\onecolumn
\appendix
\section{Halo-Model Expressions with Substructure}
We present here the full expressions for the halo-model power spectrum
and bispectrum. These expressions are derived in detail in
\citet{ShethJain2003}.

In the halo-model formalism, the power spectrum is expressed as a sum
of one- and two-halo terms:
\begin{equation}
  P(k) = P_\rmn{1h}(k) + P_\rmn{2h}(k).
\end{equation}

We add substructure by expressing the one-halo term as
\begin{equation} \label{eq:P}
  P_\rmn{1h} = P_\rmn{ss} + P_\rmn{sc} + P_\rmn{1c} + P_\rmn{2c},
\end{equation}
where
\begin{equation} \label{eq:P_ss}
  P_\rmn{ss}(k) = \int \ud M \, \frac{\ud N(M)}{\ud M}
    \left( \frac{M_\rmn{s}}{\bar\rho} \right)^2 U^2(k; M_\rmn{s})
\end{equation}
represents correlations between the smooth component of a halo. This
term looks like the one-halo term of a halo model without
substructure, but note that the mass of this smooth component is
smaller by a fraction $f$, since $M_\rmn{s} = (1 - f) M$. Hence, $P_\rmn{ss}$
is, for all $k$, less than the one-halo term calculated ignoring
substructure. It is up to the remaining substructure terms in
equation (\ref{eq:P}) to make up for this loss.
\begin{equation} \label{eq:P_sc}
  P_\rmn{sc}(k) = 2 \int \ud M \, \frac{\ud N(M)}{\ud M}
    \frac{M_\rmn{s}}{\bar\rho} \, U(k; M_\rmn{s}) \, U_\rmn{c}(k; M_\rmn{s})
    \int \ud m \,\frac{\ud n(m; M)}{\ud m}
    \, \frac{m}{\bar\rho} \, u(k; m)
\end{equation}
denotes correlations between the smooth component and a subclump. The
factor of 2 arises because either of the two density elements of the
two-point function may lie in a subclump.
\begin{equation} \label{eq:P_1c}
  P_\rmn{1c}(k) = \int \ud M \, \frac{\ud N(M)}{\ud M}
                  \int \ud m \, \frac{\ud n(m; M)}{\ud m}
                  \left( \frac{m}{\bar\rho} \right)^2 u^2(k; m)
\end{equation}
represents correlations where both elements lie in the same subclump and
\begin{equation}
  P_\rmn{2c}(k) = \int \ud M \, \frac{\ud N(M)}{\ud M} \, U_\rmn{c}^2(k; M_\rmn{s})
                  \left[ \int \ud m \, \frac{\ud n(m; M)}{\ud m}
                  \, \frac{m}{\bar\rho} \, u(k; m) \right]^2
\end{equation}
describes correlations between different subclump. Though this
expression assumes that subclumps obey a Poisson distribution, we do
not consider it important since this contribution to the power
spectrum is subdominant (see Fig.~\ref{fig:contrib}).

Finally,
\begin{equation} \label{eq:P_2h}
  P_\rmn{2h}(k) = \int \ud M_1 \, \frac{\ud N(M_1)}{\ud M_1}
                  \, \frac{M_1}{\bar\rho} \, U(k; M_1)
                  \int \ud M_2 \, \frac{\ud N(M_2)}{\ud M_2}
                  \, \frac{M_2}{\bar\rho}\, U(k; M_2)
                  \, P(k; M_1, M_2)
\end{equation}
is the two-halo term. Note that we have ignored substructure for this
contribution; on large scales we trust the approximation of smooth
NFW haloes with no substructure.

The one-halo term of the bispectrum has a similar division:
\begin{equation}
  B_\rmn{1h} = B_\rmn{sss} + B_\rmn{ssc} + B_\rmn{s1c}
               + B_\rmn{s2c} + B_\rmn{1c} + B_\rmn{2c}
               + B_\rmn{3c}.
\end{equation}
\begin{equation}
  B_\rmn{sss}(k_1, k_2, k_3) =
    \int \ud M \, \frac{\ud N(M)}{\ud M}
    \left( \frac{M_\rmn{s}}{\bar\rho} \right)^3
      U(k_1; M_\rmn{s}) \, U(k_2; M_\rmn{s}) \, U(k_3; M_\rmn{s})
\end{equation}
is the contribution with all three density elements, i.e. all three vertices
of the triangle, in the smooth component of a halo.
\begin{equation} \label{eq:B_sss} \begin{split}
    B_\rmn{ssc}(k_1, k_2, k_3) =&
    \int \ud M \, \frac{\ud N(M)}{\ud M}
    \left( \frac{M_\rmn{s}}{\bar\rho} \right)^2
        U(k_1; M_\rmn{s}) \, U(k_2; M_\rmn{s}) \, U_\rmn{c}(k_3; M_\rmn{s})\\
        & \times \int \ud m \, \frac{\ud n(m;M)}{\ud m}
        \, \frac{m}{\bar \rho} \, u(k_3; m) \\
        & + \rmn{cyclic} \ k_i \ \rmn{permutations}
\end{split} \end{equation}
places only one vertex of the triangle in a subclump.
\begin{equation} \begin{split}
    B_\rmn{s1c}(k_1, k_2, k_3) =&
            \int \ud M \,\frac{\ud N(M)}{\ud M}
            \frac{M_\rmn{s}}{\bar\rho} \, U(k_1; M_\rmn{s}) \,
            U_\rmn{c}(k_1; M_\rmn{s})\\
        &\times \int \ud m \, \frac{\ud n(m; M)}{\ud m}
            \left( \frac{m}{\bar \rho} \right)^2
            u(k_2; m) \, u(k_3; m) \\
        &+ \rmn{cyclic} \ k_i \ \rmn{permutations}
\end{split} \end{equation}
has one vertex in the smooth component and the other two in the same subclump.
\begin{equation} \begin{split}
    B_\rmn{s2c}(k_1, k_2, k_3) =&
            \int \ud M \, \frac{\ud N(M)}{\ud M}
            \frac{M_\rmn{s}}{\bar\rho} \, U(k_1; M_\rmn{s}\\
        &\times \prod_{i = 2}^3 U_\rmn{c}(k_i; M_\rmn{s})
            \int \ud m_i \, \frac{\ud n(m_i; M)}{\ud m}
            \, \frac{m_i}{\bar\rho} \, u(k_i; m_i)\\
        &+ \rmn{cyclic} \ k_i \ \rmn{permutations}
\end{split} \end{equation}
has one vertex in the smooth component and the other two in different
subclumps.
\begin{equation}
  B_\rmn{1c}(k_1, k_2, k_3) =
    \int \ud M \, \frac{\ud N(M)}{\ud M}
    \int \ud m \, \frac{\ud n(m; M)}{\ud m}
    \left( \frac{m}{\bar\rho} \right)^3
    u(k_1; m) \, u(k_2; m) \, u(k_3; m)
\end{equation}
places all three vertices in the same subclump.
\begin{equation} \begin{split}
    B_\rmn{2c}(k_1, k_2, k_3) =&
            \int \ud M \, \frac{\ud N(M)}{\ud M}
            \, U_\rmn{c}^2(k_1; M_\rmn{s}) \int \ud m_1
            \, \frac{\ud n(m_1; M)}{\ud m_1}
            \, \frac{m_1}{\bar\rho} \, u(k_1; m_1)\\
        &\times \int \ud m_2 \, \frac{\ud n(m_2; M)}
            {\ud m_2} \left( \frac{m_2}{\bar\rho} \right)^2
            u(k_2; m_2) \, u(k_3; m_3)\\
        &+ \rmn{cyclic} \ k_i \ \rmn{permutations}
\end{split} \end{equation}
divides the three vertices between two subclumps.
\begin{equation} \label{eq:B_3c}
  B_\rmn{3c}(k_1, k_2, k_3) =
    \int \ud M \, \frac{\ud N(M)}{\ud M}
    \prod_{i = 1}^3 U_\rmn{c}(k_i; M_\rmn{s}) \int \ud m_i
    \, \frac{\ud n(m_i; M)}{\ud m_i} \, \frac{m_i}{\bar\rho}
    \, u(k_i; m_i)
\end{equation}
places the three vertices in different subclumps.

\bsp
\label{lastpage}
\end{document}